# The Large Observatory For x-ray Timing

M. Feroci[1,1b]; J.W. den Herder[2]; E. Bozzo[3]; D. Barret[8]; S. Brandt[19]; M. Hernanz[6]; M. van der Klis[5]; M. Pohl[25]; A. Santangelo[17]; L. Stella[27]; A. Watts[5]; J. Wilms[87]; S. Zane[29]; M. Ahangarianabhari[14]; C. Albertus[137]; M. Alford[139]; A. Alpar[4]; D. Altamirano[5]; L. Alvarez[6]; L. Amati[7]; C. Amoros[8]; N. Andersson[9]; A. Antonelli[10]; A. Argan[1]; R. Artigue[8]; B. Artigues[6]; J.-L. Atteia[8]; P. Azzarello[3]; P. Bakala[86]; G. Baldazzi[103], S. Balman[11]; M. Barbera[12,106]; C. van Baren[2]; S. Bhattacharyya[132]; A. Baykal[11]; T. Belloni[13]; F. Bernardini[111]; G. Bertuccio[14]; S. Bianchi[59]; A. Bianchini[15]; P. Binko[3]; P. Blay[125]; F. Bocchino[99]; P. Bodin[114]; I. Bombaci[117]; J.-M. Bonnet Bidaud[16]; S. Boutloukos[17]; L. Bradley[29]; J. Braga[18]; E. Brown[49]; N. Bucciantini[21]; L. Burderi[22]; M. Burgay[115]; M. Bursa[23]; C. Budtz-Jørgensen[19]; E. Cackett[111]; F.R. Cadoux[25]; P. Cais[26]; G.A. Caliandro[6]; R. Campana[1,1b]; S. Campana[13]; F. Capitanio[1]; J. Casares[82]; P. Casella[27]; A.J. Castro-Tirado[116]; E. Cavazzuti[10]; P. Cerda-Duran[125]; D. Chakrabarty[28]; F. Château[16]; J. Chenevez[19]; J. Coker[29]; R. Cole[29]; A. Collura[106], R. Cornelisse[82]; T. Courvoisier[3]; A. Cros[8]; A. Cumming[31]; G. Cusumano[57], A. D'Aì[12]; V. D'Elia[10]; E. Del Monte[1,1b]; A. De Luca[85]; D. De Martino[31]; J.P.C. Dercksen[2]; M. De Pasquale[29]; A. De Rosa[1]; M. Del Santo[1]; S. Di Cosimo[1]; S. Diebold[17]; T. Di Salvo[12]; I. Donnarumma[1]; A. Drago[32]; M. Durant[33]; D. Emmanoulopoulos[107]; M.H. Erkut[135]; P. Esposito[85]; Y. Evangelista[1,1b]; A. Fabian[24]; M. Falanga[34]; Y. Favre[25]; C. Feldman[35]; V. Ferrari[128]; C. Ferrigno[3]; M. Finger[133]; M. H. Finger[36]; G.W. Fraser[35, +]; M. Frericks[2]; F. Fuschino[7]; M. Gabler[125]; D.K. Galloway[37]; J.L. Galvez Sanchez[6]; E. Garcia-Berro[6]; B. Gendre[10]; S. Gezari[62]; A.B. Giles[39]; M. Gilfanov[40]; P. Giommi[10]; G. Giovannini[102]; M. Giroletti[102]; E. Gogus[4]; A. Goldwurm[105]; K. Goluchová[86]; D. Götz[16]; C. Gouiffes[16]; M. Grassi[56]; P. Groot[42]; M. Gschwender[17]; L. Gualtieri[128]; C. Guidorzi[32]; L. Guy[3]; D. Haas[2]; P. Haensel[50]; M. Hailey[29]; F. Hansen[19]; D.H. Hartmann[42]; C.A. Haswell[43]; K. Hebeler[88], A. Heger[37]; W. Hermsen[2]; J. Homan[28]; A. Hornstrup[19]; R. Hudec[23,72]; J. Huovelin[45]; A. Ingram[5]; J.J.M. in't Zand[2]; G. Israel[27]; K. Iwasawa[20]; L. Izzo[47]; H.M. Jacobs[2]; F. Jetter[17]; T. Johannsen[118,127]; H.M. Jacobs[2]; P. Jonker[2]; J. Josè[126]; P. Kaaret[49]; G. Kanbach[123]; V. Karas[23]; D. Karelin[6]; D. Kataria[29]; L. Keek[49]; T. Kennedy[29]; D. Klochkov[17]; W. Kluzniak[50]; K. Kokkotas[17]; S. Korpela[45]; C. Kouveliotou[51]; I. Kreykenbohm[87]; L.M. Kuiper[2]; I. Kuvvetli[19]; C. Labanti[7]; D. Lai[52]; F.K. Lamb[53]; P.P. Laubert[2]; F. Lebrun[105]; D. Lin[8]; D. Linder[29]; G. Lodato[54]; F. Longo[55]; N. Lund[19]; T.J. Maccarone[131]; D. Macera[14]; S. Maestre[8]; S. Mahmoodifar[62]; D. Maier[17]; P. Malcovati[56]; I. Mandel[120]; V. Mangano[144]; A. Manousakis[50]; M. Marisaldi[7]; A. Markowitz[109]; A. Martindale[35]; G. Matt[59]; I.M. McHardy[107]; A. Melatos[60]; M. Mendez[61]; S. Mereghetti[85]; M. Michalska[68], S. Migliari[20]; R. Mignani[85,108]; M.C. Miller[62]; J.M. Miller[49]; T. Mineo[57]; G. Miniutti[112]; S. Morsink[64]; C. Motch[65]; S. Motta[13]; M. Mouchet[66]; G. Mouret[8]; J. Mulačová[19]; F. Muleri[1,1b]; T. Muñoz-Darias[140]; I. Negueruela[95]; J. Neilsen[28]; A.J. Norton[43]; M. Nowak[28]; P. O'Brien[35]; P.E.H. Olsen[19]; M. Orienti[102]; M. Orio[99,110]; M. Orlandini[7]; P. Orleanski[68]; J.P. Osborne[35]; R. Osten[69]; F. Ozel[70]; L. Pacciani[1,1b]; M. Paolillo[119]; A. Papitto[6]; J. M. Paredes[20]; A. Patruno[83,141]; B. Paul[71]; E. Perinati[17]; A. Pellizzoni[115]; A.V. Penacchioni[47]; M.A. Perez[136]; V. Petracek[72]; C. Pittori[10]; J. Pons[95]; J. Portell[6]; A. Possenti[115]; J. Poutanen[73]; M. Prakash[122]; P. Le Provost[16]; D. Psaltis[70]; D. Rambaud[8]; P. Ramon[8]; G. Ramsay[76]; M. Rapisarda[1,1b]; A. Rachevski[77]; I. Rashevskaya[77]; P.S. Ray[78]; N. Rea[6]; S. Reddy[80]; P. Reig[113,81]; M. Reina Aranda[63]; R. Remillard[28]; C. Reynolds[62]; L. Rezzolla[124]; M. Ribo[20]; R. de la Rie[2]; A. Riggio[115]; A. Rios[138]; P. Rodríguez-Gil[82,104]; J. Rodriguez[16]; R. Rohlfs[3]; P. Romano[57]; E.M.R. Rossi[83]; A. Rozanska[50]; A. Rousseau[29]; F. Ryde[84]; L. Sabau-Graziati[63]; G. Sala[6]; R. Salvaterra[85]; A. Sanna[61]; J. Sandberg[134]; S. Scaringi[130]; S. Schanne[16]; J. Schee[86]; C. Schmid[87]; S. Shore[117]; R. Schneider[27]; A. Schwenk[88]; A.D. Schwope[89];

J.-Y. Seyler[114]; A. Shearer[90]; A. Smith[29]; D.M. Smith[58]; P.J. Smith[29]; V. Sochora[23]; P. Soffitta[1]; P. Soleri[61]; A. Spencer[29]; B. Stappers[91]; A.W. Steiner[80]; N. Stergioulas[92]; G. Stratta[10]; T.E. Strohmayer[93]; Z. Stuchlik[86]; S. Suchy[17]; V. Sulemainov[17]; T. Takahashi[94]; F. Tamburini[15]; T. Tauris[129]; C. Tenzer[17]; L. Tolos[6]; F. Tombesi[62]; J. Tomsick[121]; G. Torok[86]; J.M. Torrejon[95]; D.F. Torres[96]; A. Tramacere[3]; A. Trois[1]; R. Turolla[15]; S. Turriziani[101], P. Uter[17]; P. Uttley[5]; A. Vacchi[77]; P. Varniere[105]; S. Vaughan[35]; S. Vercellone[57]; V. Vrba[97]; D. Walton[29]; S. Watanabe[94]; R. Wawrzaszek[68]; N. Webb[8]; N. Weinberg[28]; H. Wende[17]; P. Wheatley[98]; R. Wijers[5]; R. Wijnands[5]; M. Wille[87]; C.A. Wilson-Hodge[44]; B. Winter[29]; K. Wood[78]; G. Zampa[77]; N. Zampa[77]; L. Zampieri[99]; L. Zdunik[50]; A. Zdziarski[50]; B. Zhang[100]; F. Zwart[2], M. Ayre[142], T. Boenke[142], C. Corral van Damme[142], E. Kuulkers[143], D. Lumb[142].

[1]IAPS-INAF, Via del Fosso del Cavaliere 100 - 00133 Rome, Italy; [1b]INFN, Sez. Roma Tor Vergata, Via della Ricerca Scientifica 1 - 00133 Rome, Italy; [2]SRON, Sorbonnelaan 2 - 3584 CA Utrecht, The Netherlands; [3]ISDC, Geneve University, Chemin d'Ecogia 16 - 1290 Versoix, Switzerland; [4]Sabanci University, Orhanli-Tuzla 34956, Istanbul, Turkey; [5]Astronomical Institute Anton Pannekoek, University of Amsterdam, Science Park 904 - 1098 XH Amsterdam, The Netherlands; [6]IEEC-CSIC-UPC-UB, Carrer del Gran Capità, 2 - 08034 Barcelona, Spain; [7]INAF-IASF-Bologna, Via P. Gobetti, 101 - 40129 Bologna, Italy; [8]IRAP, avenue du Colonel Roche, 9 - BP 44346 Toulouse, France; [9]Faculty of Physical and Applied Sciences, University of Southampton, Southampton, SO17 1BJ, United Kingdom; [10]ASDC, Via del Politecnico snc - 00133 Rome, Italy; [11]Middle East Technical University, Ankara, Mah. Dumlupınar Blv. No:1 - 06800 Çankaya Ankara, Turkey; [12]Dipartimento di Chimica e Fisica, Palermo University, Via Archirafi, 36 - 90123 Palermo, Italy; [13]INAF-OA Brera, Via E. Bianchi 46 – 23807 Merate (LC), Italy; [14]Politecnico Milano, Piazza Leonardo da Vinci, 32 - 20133 Milano, Italy; [15]Dept. of Physics and Astronomy University of Padua, vicolo Osservatorio 3, 35122, Padova, Italy; [16]CEA Saclay, DSM/IRFU/SAp, 91191 Gif sur Yvette, France; [17] IAAT University of Tuebingen, Sand 1 - 72076 Tuebingen, Germany; [18]INPE, Avenida dos Astronautas 1.758, Jd. da Granja - 12227-010 São José dos Campos, Brazil; [19]National Space Institute, Technical University of Denmark, Elektrovej Bld 327, 2800 Kgs Lyngby, Denmark; [20]DAM and ICC-UB, Universitat de Barcelona (IEEC-UB), Martì i Franquès 1, E-08028, Barcelona, Spain; [21]Arcetri Observatory, INAF, Largo Enrico Fermi 5 - I-50125 Firenze, Italy; [22]Cagliari University, Strada provinciale per Sestu, KM 1 - 09042 Monserrato, Italy; [23]Astronomical Institute of the Academy of Sciences of the Czech Republic, Fricova 298, CZ-251 65 Ondrejov, Czech Republic; [24]Cambridge University, Trinity Lane - CB2 1TN Cambridge, United Kingdom; [25]DPNC, Geneve University, Quai Ernest-Ansermet 30 - 1205 Geneva, Switzerland; [26]Laboratoire d'Astrophysique de Bordeaux, rue de l'Observatoire - BP 89 - 33270 Floirac Cedex, France; [27]INAF-OA Roma, Via Frascati, 33 - 00040 Monte Porzio Catone, Italy; [28]MIT, 77 Massachusetts Avenue - MA 02139 Cambridge, United States; [29]MSSL, Holmbury St Mary - RH5 6NT Dorking, Surrey, United Kingdom; [30]McGill University, 845 Sherbrooke Street West - QC H3A 0G4 Montréal, Canada; [31]INAF-OA Capodimonte, Salita Moiariello, 16 - 80131 Napoli, Italy; [32]Ferrara University, Via Saragat 1 - 44122 Ferrara, Italy; [33]Department of Medical Biophysics, University of Toronto, M4N 3M5 Canada; [34]ISSI Bern, Hallerstrasse 6 - 3012 Bern, Switzerland; [35]Leicester University, University Road - LE1 7RH Leicester, United Kingdom; [36]Universities Space Research Association, 6767 Old Madison Pike, Suite 450, Huntsville, Alabama 35806, United States; [37] Monash Centre for Astrophysics, School of Physics and School of Mathematical Sciences, Monash University, Clayton VIC 3800, Australia; [38]Johns Hopkins University, 3400 North Charles Street - Baltimore, United States; [39] University of Tasmania, Private Bag 37, Hobart, TAS 7001, Australia; [40]MPA Garching, Karl-Schwarzschild-Str. 1 - 85741 Garching, Germany; [41]Radboud University, Heyendaalseweg 135 - 6500 GL Nijmegen, The Netherlands; [42]Clemson University, Clemson, SC 29634, United States; [43]Open University, Walton Hall - MK7 6AA Milton Keynes, United Kingdom; [44]Astrophysics Office, ZP12, NASA/Marshall Space Flight Center, Huntsville, AL 35812, United States; [45]Department of Physics, Division of Geophysics and Astronomy, P.O. Box 48, FI-00014 University of

Helsinki, Finland; [46]Durham University, Stockton Rd - DH1 3UP Durham, United Kingdom; [47]Sapienza University and ICRA, p.le A. Moro, 2 - 00185, Rome, Italy; [48]University of Iowa, Van Allen Hall, Iowa City, IA 52242, United States; [49]Michigan state University, 567 Wilson Rd, - MI 48824 East Lansing, United States; [50]Copernicus Astronomical Center, Bartycka 18 - Warsaw, Poland; [51]Science and Technology Office, ZP12, NASA/Marshall Space Flight Center, Huntsville, AL 35812, United States; [52]Cornell University, Space Building - NY 14853 Ithaca, United States; [53] University of Illinois, Physics Department, 1110 W. Green St., Urbana, IL 61801, United States; [54]Dipartimento di Fisica, Università degli Studi di Milano, Via Celoria 16, 20133 Milano, Italy; [55]University of Trieste, Via Alfonso Valerio, 32 - 34128 Trieste, Italy; [56]Pavia University, Corso Strada Nuova, 65 - 27100 Pavia, Italy; [57]INAF IFC, Via Ugo La Malfa, 153 - 90146 Palermo, Italy; [58]University of California, Santa Cruz, 1156 High St. CA 95064 Santa Cruz, United States; [59]University of Rome III, Via della Vasca Navale, 84 - 00146 Roma, Italy; [60]University of Melbourne, Swanston Street - VIC 3052 Parkville, Australia; [61]Kapteyn Astronomical Institute, University of Groningen, P.O. Box 800, 9700 AV Groningen, The Netherlands; [62]University of Maryland, Department of Astronomy, College Park, MD 20742-2421, United States; [63]National Institute of Aerospace Technology (INTA), Carretera de Ajalvir km. 4 - 28850 Torrejón de Ardoz, Spain; [64]University of Alberta, 85 Avenue 116 St NW - AB T6G 2R3 Edmonton, Canada; [65]Observatoire Astronomique de Strasbourg, 11 rue de l'Université - 67000 Strasbourg, France; [66]Université Paris Diderot 5 rue Thomas-Mann 75205 Paris cedex 13, France; [67]INAF-OA Torino, Via Osservatorio, 20 - 10025 Pino Torinese, Italy; [68]Space Research Centre, Warsaw, Bartycka 18A - Warszawa, Poland; [69]Space Telescope Institute, 3700 San Martin Drive - MD 21218 Baltimore, United States; [70]University of Arizona, Department of Astronomy, 933 N. Cherry Ave, Tucson, AZ 85721, United States; [71]Raman Research Institute, C. V. Raman Avenue - 560 080 Sadashivanagar, India; [72]Czech Technical University in Prague, Zikova 1903/4, CZ-166 36 Praha 6, Czech Republic; [73]Tuorla Observatory, University of Turku, Vaisalantie 20, FI-21500 Piikkio, Finland; [76]Armagh Observatory, College Hill - BT61 9DG Armagh, United Kingdom; [77]INFN, Trieste, Via A. Valerio 2 - I-34127 Trieste, Italy; [78]NRL, 4555 Overlook Ave. SW Washington, DC 20375-5352, United States; [80]Institute for Nuclear Theory, University of Washington, WA 98195-1550 - Seattle, United States; [81]Physics Department, University of Crete, GR-710 03 Heraklion, Greece; [82]Instituto de Astrofisica de Canarias, Vía Láctea s/n, La Laguna, E-38205, Tenerife, Spain; [83]Leiden Observatory, Niels Bohrweg 2 - NL-2333 CA Leiden, The Netherlands; [84]KTH Royal Institute of Technology, Valhallavägen 79 - 100 44 Stockholm, Sweden; [85]INAF-IASF-Milano, Via E. Bassini 15 - I-20133 Milano, Italy; [86]Silesian University in Opava, Na Rybníčku 626/1 - 746 01 Opava, Czech Republic; [87]University of Erlangen-Nuremberg, Schlossplatz 4 - 91054 Erlangen, Germany; [88]Institut für Kernphysik, Technische Universität Darmstadt, 64289 Darmstadt, Germany and ExtreMe Matter Institute EMMI, GSI Helmholtzzentrum für Schwerionenforschung GmbH, 64291 Darmstadt; [89]Leibniz-Institut fuer Astrophysik Potsdam, An der Sternwarte 16 - 14482 Potsdam, Germany; [90]National University of Ireland Galway, University Road, Galway, Ireland; [91]University of Manchester, Booth Street West - M15 6PB Manchester, United Kingdom; [92]Aristotle University of Thessaloniki, Greece; [93]Goddard Space Flight Center, 8800 Greenbelt Rd. - Md., 20771 Greenbelt, United States; [94]ISAS, 3-1-1 Yoshinodai, Chuo-ku, Sagamihara - 252-5210 Kanagawa, Japan; [95]University of Alicante, Carretera San Vicente del Raspeig - 03690 Sant Vicent del Raspeig, Spain; [96]ICREA - Institució Catalana de Recerca i Estudis Avançats, Passeig Lluís Companys, 23 - 08010 Barcelona, Spain; [97]Physical Institute of the Academy of Sciences of the Czech Republic, Na Slovance 1999/2, CZ-182 21 Praha 8, Czech Republic; [98]University of Warwick, Gibbet Hill Road - CV4 7AL Coventry, United Kingdom; [99]INAF-OA Padova, Vicolo Osservatorio 5, Padova, Italy; [100]University of Nevada, Las Vegas, NV 89012, United States; [101]University of Rome Tor Vergata, Via della Ricerca Scientifica 1 - 00133 Rome, Italy; [102]INAF-IRA-Bologna, Via P. Gobetti 101- 40129 Bologna, Italy; [103]University of Bologna, Dept. of Physics and INFN section of Bologna, V.le Berti Pichat, 6/2, 40127, Bologna, Italy; [104]Departamento de Astrofisica, Universidad de La Laguna, La Laguna, E-38206, Santa Cruz de Tenerife, Spain; [105]APC, AstroParticule & Cosmologie, UMR 7164 CNRS/N2P3, Université Paris Diderot, CEA/Irfu, Observatoire de Paris, Sorbonne Paris Cite 10 rue Alice Domon et Leonie Duquet, 75205 Paris Cedex 13, France; [106]INAF- Osservatorio Astronomico di Palermo, Piazza del Parlamento 1, 90134 Palermo,

Italy; [107]School of Physics and Astronomy, University of Southampton, Southampton, SO17 1BJ, UK; [108]Kepler Institute of Astronomy, University of Zielona Gòra, Lubuska 2, 65-265, Zielona Gòra, Poland; [109]University of California, San Diego, Mail Code 0424, La Jolla, CA 92093 United States; [110]Dept. of Astronomy, Univ. of Wisconsin, 475 N. Charter Str., Madison WI 53706, United States; [111]Wayne State University, Department of Physics & Astronomy, 666 W. Hancock St, Detroit, MI 48201, United States; [112]Centro de Astrobiologia (CSIC--INTA), P.O. Box 78, E-28691, Villanueva de la Cañada, Madrid, Spain; [113]Foundation for Research and Technology - Hellas, GR-711 10 Heraklion, Greece; [114]CNES, 18 Avenue Edouard Belin, 31400 Toulouse, France; [115]INAF-OA Cagliari, località Poggio dei Pini, Strada 54, 09012 Capoterra, Italy; [116]Instituto Astrofisica de Andalucia, Glorieta de la Astronomía (IAAC-CSIC), s/n., E-18008, Granada, Spain; [117]University of Pisa, Largo B. Pontecorvo 3, Pisa, I-56127 Italy; [118]Perimeter Institute for Theoretical Physics, 31 Caroline Street North, Waterloo, ON, N2L 2Y5, Canada; [119]Dipartimento di Scienze Fisiche, Università di Napoli Fedelico II, C.U. di Monte Sant'Angelo, Via Cintia ed. 6, 80126, Napoli, Italy; [120]School of Physics and Astronomy, University of Birmingham, Edgbaston B15 2TT, Birmingham, United Kingdom; [121]University of California, Berkeley, Space Sciences Laboratory, 7 Gauss Way, Berkeley, CA 94720-7450; [122]Ohio University, Department of Physics & Astronomy, Athens, OH 45701, United States; [123]Max-Planck-Institut fuer extraterrestrische Physik, Postfach 1603, 85740 Garching, Germany; [124]Max Planck Institute for Gravitational Physics (Albert Einstein Institute), Am Mühlenberg 1, D-14476 Golm, Germany; [125]University of Valencia, Av de Vicente Blasco Ibáñez, 13, 46010 Valencia, Spain; [126]Technical University of Catalonia, C. Jordi Girona, 31, 08034 Barcelona, Spain; [127]Department of Physics and Astronomy, University of Waterloo, 200 University Avenue West, Waterloo, ON, N2L 3G1, Canada; ; [128]Sapienza University, p.le A. Moro, 2 - 00185, Rome, Italy; [129]Argelander-Institut für Astronomie, Auf dem Hügel 71, D-53121, Bonn, Germany; [130]Institute for Astronomy K.U. Leuven, Celestijnenlaan 200D, 3001, Leuven Belgium; [131]Texas Tech University, Physics Department, Box 41051, Lubbock, TX 79409-1051; [132]Tata Institute of Fundamental Research, 1 Homi Bhabha Road, Colaba, Mumbai 400005, India; [133]Charles University in Prague, Faculty of Mathematics and Physics, V Holesovickach 2, CZ18000 Prague, Czech Republic; [134]Jorgen Sandberg Consulting, Denmark; [135]Istanbul Kültür University, Faculty of Science and Letters, Ataköy Campus, Bakırköy 34156, Istanbul, Turkey; [136]Fundamental Physics Department, Facultad de Ciencias-Trilingüe University of Salamanca, Plaza de la Merced s/n, E-37008 Salamanca, Spain; [137]Departamento de Física Atómica, Molecular y Nuclear, Facultad de Ciencias, Universidad de Granada, E-18071-Granada-Spain; [138]University of Surrey, Guildford, Surrey GU2 7XH, United Kingdom; [139] Washington University, Dept. of Physics- Compton Hall, One Brookings Drive - Campus Box 1105, St. Louis, MO 63130, United States; [140]Oxford University, Department of Physics, Clarendon Laboratory, Parks Road, Oxford, OX1 3PU, United Kingdom; [141]ASTRON, the Netherlands Institute for Radio Astronomy, Postbus 2, 7990 AA Dwingeloo, The Netherlands; [142]European Space Agency, ESTEC, Keplerlaan 1, 2201 AZ Noordwijk, The Netherlands; [143]European Space Astronomy Centre, SRE-O, PO Box 78, 28691 Villanueva de la Cañada, Madrid, Spain; [144]Department of Astronomy and Astrophysics, The Pennsylvania State University, 525 Davey Lab, University Park, PA 16802, USA

[*] marco.feroci@inaf.it

[+] This work is dedicated to the memory of Prof. George Fraser, a dear friend and esteemed colleague.

## ABSTRACT

The Large Observatory For x-ray Timing (LOFT) was studied within ESA M3 Cosmic Vision framework and participated in the final downselection for a launch slot in 2022-2024. Thanks to the unprecedented combination of effective area and spectral resolution of its main instrument, LOFT will study the behaviour of matter under extreme conditions, such as the strong gravitational field in the innermost regions of accretion flows close to black holes and neutron stars, and the supranuclear densities in the interior of neutron stars. The science payload is based on a Large Area Detector (LAD, 10 $m^2$ effective area, 2-30 keV, 240 eV spectral resolution, 1° collimated field of view) and a Wide

Field Monitor (WFM, 2-50 keV, 4 steradian field of view, 1 arcmin source location accuracy, 300 eV spectral resolution). The WFM is equipped with an on-board system for bright events (e.g. GRB) localization. The trigger time and position of these events are broadcast to the ground within 30 s from discovery. In this paper we present the status of the mission at the end of its Phase A study.

## 1. INTRODUCTION

LOFT is designed to observe the rapid spectral and flux variability of X-rays emitted from regions close to the surface of neutron stars and the event horizon of black holes. The proposed measurements are efficient diagnostics of the behavior and motion of matter in the presence of strong gravitational fields, where the effects predicted by General Relativity are largest, and of the physics of matter at densities in excess of that in atomic nuclei, determining its equation of state and composition. This research addresses science theme 3 proposed in the ESA Cosmic Vision programme: “What are the fundamental physical laws of the Universe”. The scientific payload was studied by a consortium of European scientific institutes, including teams from the Czech Republic, Denmark, Finland, France, Germany, Italy, the Netherlands, Poland, Spain, Switzerland and the United Kingdom, with support from international partners in Brazil, India, Japan and the United States. An even wider science support community contributed scientific inputs to help focus and refine the science case and scientific requirements.

The scientific payload of the LOFT mission includes two experiments (see Figure 1): the Large Area Detector (LAD, [1]) and the Wide Field Monitor (WFM, [2]). The key feature of the LAD is its very large area combined with “CCD-class” energy resolution. The 1-deg collimated field of view LAD will be able to access ≥75% of the sky at any time, to observe Galactic and bright extragalactic sources in their most interesting states. To achieve this, LOFT is equipped with the WFM, which will monitor more than half of the LAD-accessible sky (approximately 1/3 of the whole sky) simultaneously at any time. The WFM operates in the same energy range as the LAD, providing information about source state (flux variability and energy spectrum), as well as arc-minute positioning. With such a wide angle sky monitoring, the WFM will also provide long-term histories of the target sources, serving to enhance and provide context to the LAD observations as well as enabling a series of wider science goals.

The combination of enormous effective area, improving by a factor of 15 over the largest predecessor (RossiXTE/PCA), with good spectral resolution, improving by a factor >100 in throughput over the largest predecessor XMM-Newton, makes LOFT a unique breakthrough observing facility in X-ray astronomy, uniting for the first time two classical observing “routes” in X-ray astronomy, timing and spectroscopy. This will enable not only to achieve the design science goals, but will open an enormous discovery space for many classes of high energy sources.

The LOFT mission concept (Large Observatory For x-ray Timing, [3]) was submitted on December 2010 in response to the M3 call issued by the European Space Agency (ESA) within the framework of the Cosmic Vision 2015-2025 programme. The mission was selected as one of the four M3 candidate missions and underwent its Assessment Study in 2011-2013. In February 2014 ESA selected the M3 mission. Despite excellent evaluation of the science and a sound technical and programmatic assessment within the M3 boundary conditions, LOFT was not eventually selected as the mission to be implemented in this context.

In this paper we provide an overview of the LOFT science drivers and goals, a short description of the payload, ground segment and mission profile. More detailed descriptions of the LAD and WFM instruments may be found in Zane et al. [1] and Brandt et al. [2], respectively, while the science ground segment is described in Bozzo et al. [4].

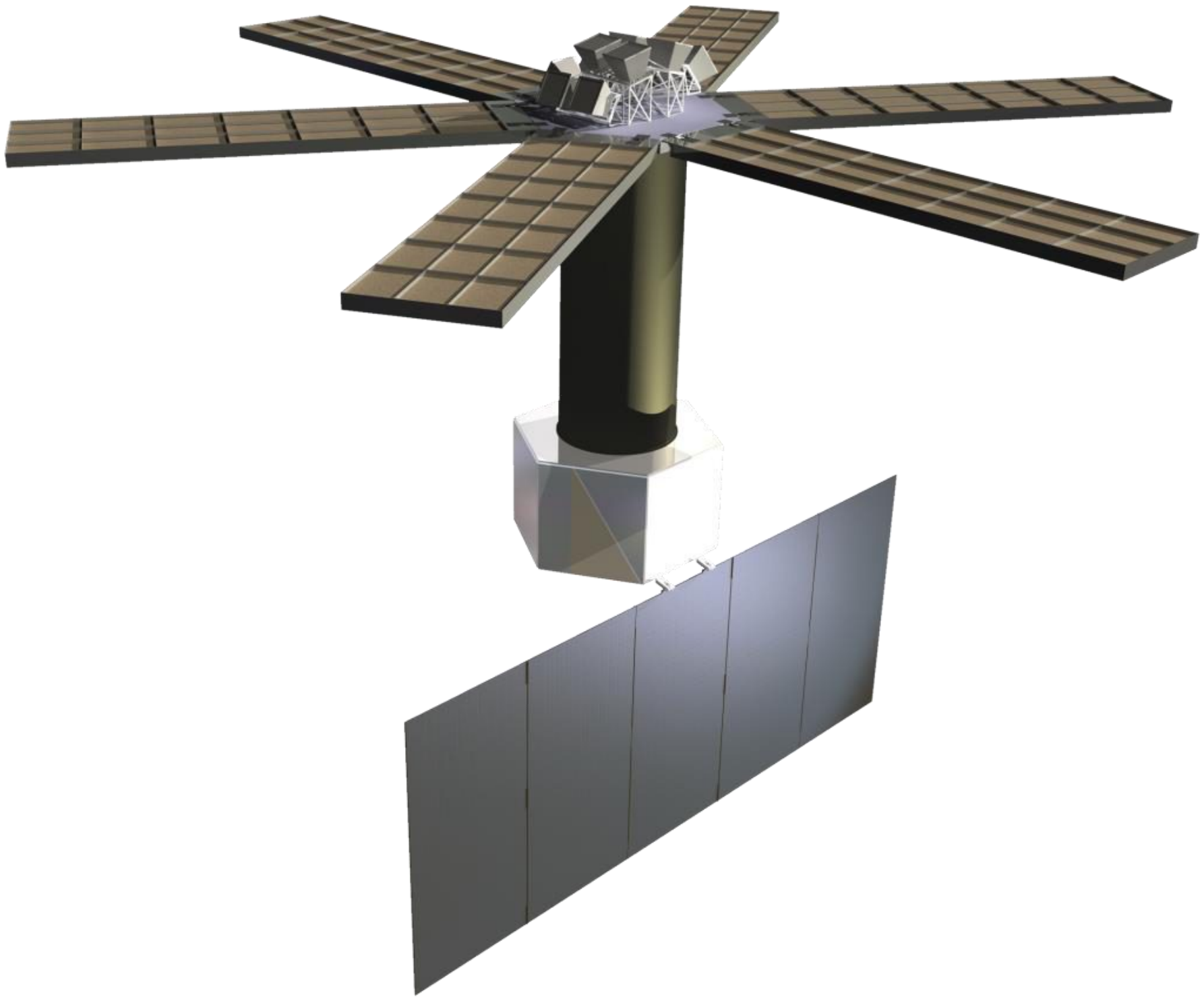

Figure 1 The LOFT satellite (Consortium configuration).

## 2. SCIENCE OBJECTIVES

The core science goals of the LOFT mission are the study of matter in ultradense environments ("Dense Matter") and under strong gravitational fields ("Strong Field Gravity"). These two major themes drive the design of the mission. The LOFT approach is to exploit the powerful diagnostics of time and spectral variability of cosmic sources which occur on dynamical timescales (e.g., the orbital timescale of the inner accretion disk around a black hole). In addition, despite being designed for its core science goals, the unique characteristics of LOFT makes it an ideal observatory to study a wide range of celestial objects for which a large fraction of the observing time remains available. This goes under the general category of "Observatory Science".

In the following sections we briefly summarize the rationale of the LOFT approach for the main science objectives and list the quantitative top-level science goals identified by the LOFT Science Study Team. A more extended discussion on these themes will be reported in forthcoming papers by the LOFT science working groups. A concise description may already be found in the science report of the LOFT phase A study, the so-called Yellow Book.

### 2.1 Equation of state of ultradense matter

One of the most important open questions in high energy astrophysics and nuclear physics is the relation between density and pressure of matter at densities larger than that of atomic nuclei, where quantum chromo-dynamics (QCD) and quark physics are poorly constrained. While being of the highest interest, this high-density/low-temperature region of the QCD phase diagram is inaccessible to high-energy terrestrial experiments and can only be probed through the relation of density and pressure, i.e., the equation of state (EoS) of neutron stars (NS). Many models for the EoS have been proposed over the years, reflecting different possible QCD properties and phases. These allow to predict how the mass M and the radius R of NS are expected to be related, thus offering testable predictions. Recent accurate measurements of high NS masses in binary radio pulsars have provided significant constraints, but the issue of determining the equation of state remains open. This requires measuring both M and R with high precision (high statistical quality and low systematics both in measurements and models), for a wide range of NS masses. While more precise mass estimates are expected from radio observations, highly precise and accurate measurements of the NS radius and mass in the same object can be obtained through X-ray observations.

The LOFT approach to the EoS is to use multiple diagnostics to provide several (M, R) high precision measurements. They include pulse profile modelling, a more complete NS spin distribution and the measurement of global seismic oscillations during intermediate flares from magnetars. Modeling of pulse profiles is the most promising. The shape of the periodic signal in accreting millisecond spinning NS and during thermonuclear bursts encode information about M and R, as the photons propagate through curved space-time and affected by relativistic effects. Modeling of these distortions (Doppler boosting, time dilation, gravitational light bending and frame dragging), enabled by high statistics, will provide accurate measurements of M and R. This technique will be applied to oscillations during the rise and tail of bursts, as well as transient coherent signals of magnetospheric origin, offering independent measurements also on the same source. In order to reach the required understanding and thus reducing the systematics in the modeling of the phenomenon, a deep study has been carried out by the LOFT Dense Matter working group. This is reported in Lo et al. [5] and Watts et al. [6].

The key LOFT requirements with respect to determining the dense matter EOS have been expressed as 3 top-level science goals:

*EOS1* Constrain the equation of state of supranuclear-density matter by the measurement, using three complementary types of pulsations, of mass and radius of at least 4 neutron stars with an instrumental accuracy of 4% in mass and 3% in radius.

*EOS2* Provide an independent constraint on the equation of state by filling out the accreting neutron star spin distribution through discovering coherent pulsations down to an amplitude of about 0.4% (2%) rms for a 100 mCrab (10 mCrab) source in a time interval of 100 s, and oscillations during type I bursts down to typical amplitudes of 1% (2.5%) rms in the burst tail (rise) among 35 neutron stars covering a range of luminosities and inclinations.

*EOS3* Probe the interior structure of isolated neutron stars by observing seismic oscillations in Soft Gamma-ray Repeater intermediate flares when they occur with flux ~1000 Crab through high energy photons (> 20 keV).

The anticipated LOFT results on the EoS are summarized in Figure 2.

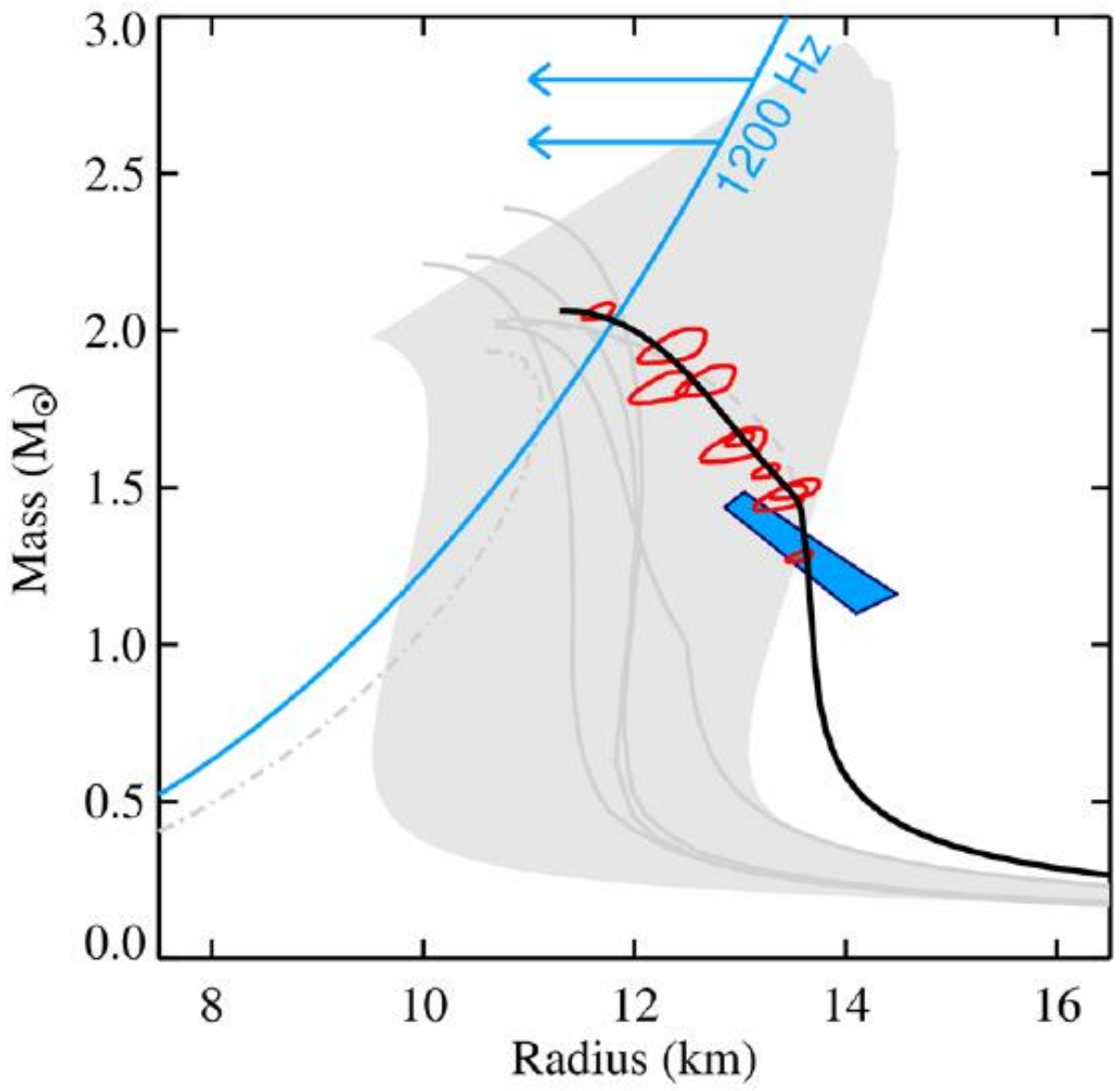

Figure 2 A summary of the LOFT measurements of the equation of state in neutron stars. Red circles derive from pulse modeling, the blue line identifies the constraint derived from the NS spin distribution, while the blue trapezoid shows the limit offered by global seismic oscillations. The combination of these independent constrains will allow to map the equation of state (see LOFT Yellow Book: http://sci.esa.int/loft/53447-loft-yellow-book/).

### 2.2 Behavior of matter under strong gravity

General Relativity (GR) has been probed with high accuracy in the weak-field regime, for gravitational radii $r_g \sim 10^5$-$10^6$ where relativistic effects are only small perturbations. X-rays emitted from the innermost regions (a few $r_g$) around compact objects (neutron stars and black holes) originate from matter experiencing gravity I the strong field regime and provide the best tools to explore the physics under such extreme conditions, testing the predictions of GR where they are expected to produce macroscopic effects. LOFT will be able to approach this science goal from different angles, using both its large area and its high spectral resolution. Black holes and neutron stars display quasi-periodic oscillations (QPOs) in their X-ray flux arising from the millisecond dynamical timescales of the inner accretion flows. The interpretation of such high frequency QPOs necessarily involves fundamental frequencies of the motion of matter orbiting in disk regions dominated by the gravitational field. Examples of such interpretations are the models attributing high frequency QPOs to relativistic radial and vertical epicyclic frequencies versus those predicting them to arise from relativistic nodal and periastron precession. The LOFT measurements will discriminate between such models and directly access so far untested GR effects, such as frame-dragging, strong-field periastron precession, and the existence of an innermost stable orbit around black holes.

LOFT will open a new era in the field of X-ray timing, providing access to information that is qualitatively new, due to its capability to measure dynamical timescale phenomena on their coherence time: previously we have been able to study only time-averaged behavior. One beautiful example serves to illustrate LOFT's technical capabilities. By combining large area with good spectral energy resolution, low frequency (e.g. ~30 Hz) QPOs will be detected in neutron star binaries at such high statistical accuracy to allow true phase-resolved spectroscopy, as for a coherent signal. The observation of the variable Fe K line profile at different phases will permit detection of the expected features of Lense-Thirring precession of the inner disk at $\sim r_g$, providing a measurement of the inclination of the varying ring.

The variability of the Fe K line profile is also a tool to measure mass and spin of black-holes. This is possible using both Galactic and extragalactic black holes, but is best done using Active Galactic Nuclei, where the longer dynamical timescales compensate for their dimmer flux, providing better counting statistics per dynamical timescale. Fe-line tomography in a few tens of bright AGN are expected to provide significant constraints on the mass and spin of their supermassive black holes.

Reverberation mapping (that is, measuring the delay between direct continuum and disk-reflected radiation, e.g. Uttley et al. [7] for a review) at the Fe-line energies provides a unique tool to measure dynamical parameters in strong field, as a function of the absolute distance to the black hole. For AGNs LOFT will improve by a factor of ~6 over XMM-Newton, but it will really open this diagnostics to the much brighter X-ray binaries, offering an increase in sensitivity by more than a factor of ~200, a real breakthrough.

The key LOFT requirements with respect to SFG have been expressed as 5 top-level science goals:

*SFG1* Detect strong-field GR effects by measuring epicyclic motions in high frequency QPOs from at least 3 black hole X-ray binaries and perform comparative studies in neutron stars.

*SFG2* Detect disk precession due to relativistic frame dragging with the Fe line variations in low frequency QPOs for 10 neutron stars and 5 black holes.

*SFG3* Detect kHz QPOs at their coherence time, measure the waveforms and quantify the distortions due to strong field GR for 10 neutron stars covering different inclinations and luminosities.

*SFG4* Constrain fundamental properties of stellar mass black holes and of accretion flows in strong field gravity by (a) measuring the Fe-line profile and (b) carrying out reverberation mapping and (c) tomography of 5 black holes in binaries providing spins to an accuracy of 5% of the maximum spin (a/M=1) and do comparative studies in 10 neutron stars

*SFG5* Constrain fundamental properties of supermassive black holes and of accretion flows in strong field gravity by (a) measuring the Fe-line profiles of 20 AGNs and for 6 AGNs (b) carry out reverberation mapping and (c) tomography, providing BH spins to an accuracy of 20% of the maximum spin (10% for fast spins) and measuring their masses with 30% accuracy.

The LOFT Science Study Team has translated the top-level goals EOS 1-3 and SFG 1-5 into a sequence of realistic observations that will allow meeting the science requirements. This involved identifying the best targets determining whether the relevant observations can be planned in advance or must be carried out as Targets of Opportunity (ToOs), the anticipated number of pointings and the total observing time necessary to reach the required science goal.

### 2.3 Observatory science

The EOS and SFG areas are the primary LOFT science drivers, meaning that they set the tightest requirements on the main instrumental properties, such as the effective area or spectral resolution. Clearly, an instrument like the LAD – with CCD-class energy resolution over a wide band, combined with 10 m$^2$ effective area – or the WFM – with soft X-ray bandpass, ~300 eV energy resolution and arcmin imaging simultaneously over ~3-4 steradians – will dramatically change the observational scenario for a wide range of both Galactic and extragalactic sources. Time variability and spectroscopy will become accessible with the LAD on as yet unexplored timescales, while the WFM will provide spectral information on unpredictable flares and bursts from Galactic sources as well as cosmic gamma-ray bursts.

An overview of the potential discovery space offered by LOFT as a general observatory cannot be given here. Below we just illustrate by example in figures the quality of science data expected from both the LAD and the WFM.

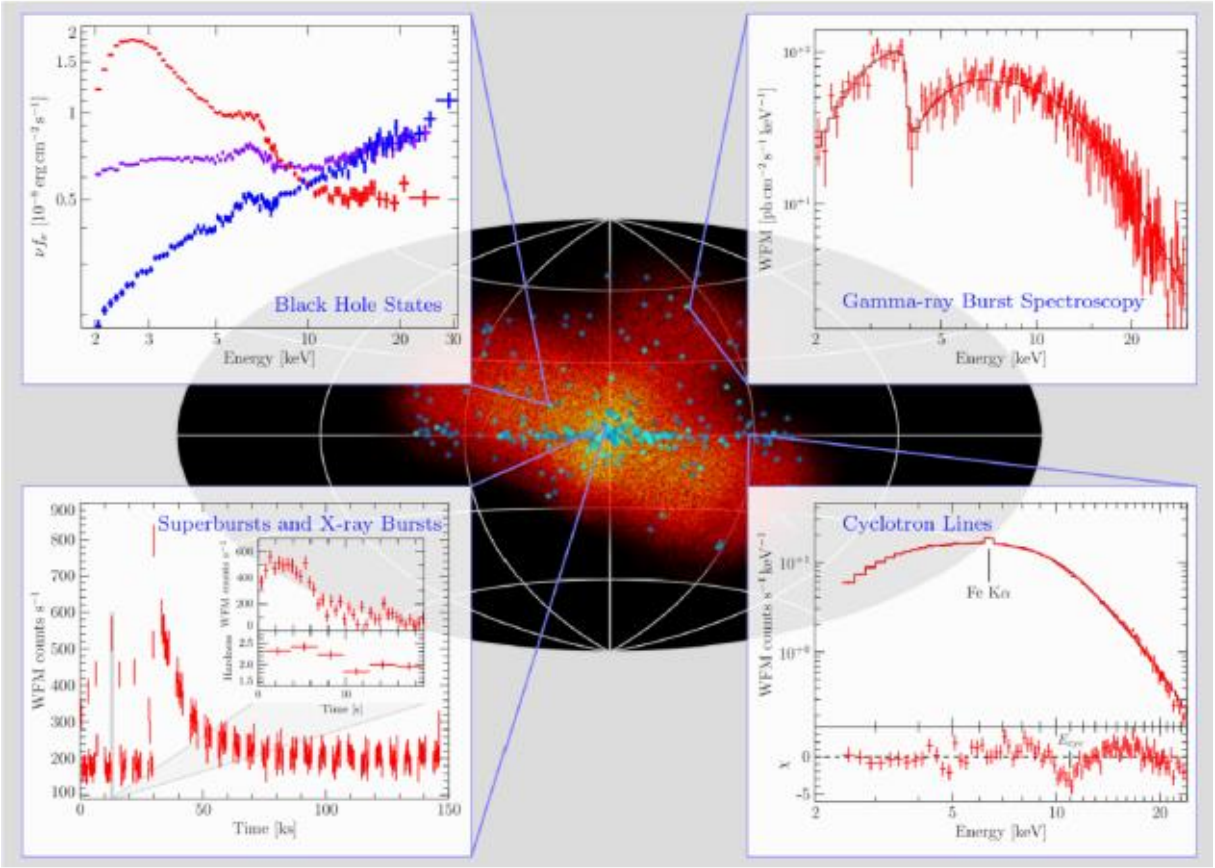

Figure 3 A simulated field observed by the Wide Field Monitor, highlighting the potential of simultaneously detecting state changes in black-hole binaries (top left insert), an absorption edge in the prompt emission of gamma-ray bursts (top right), type I X-ray bursts and superbursts (bottom left) and cyclotron lines in high mass X-ray binaries.

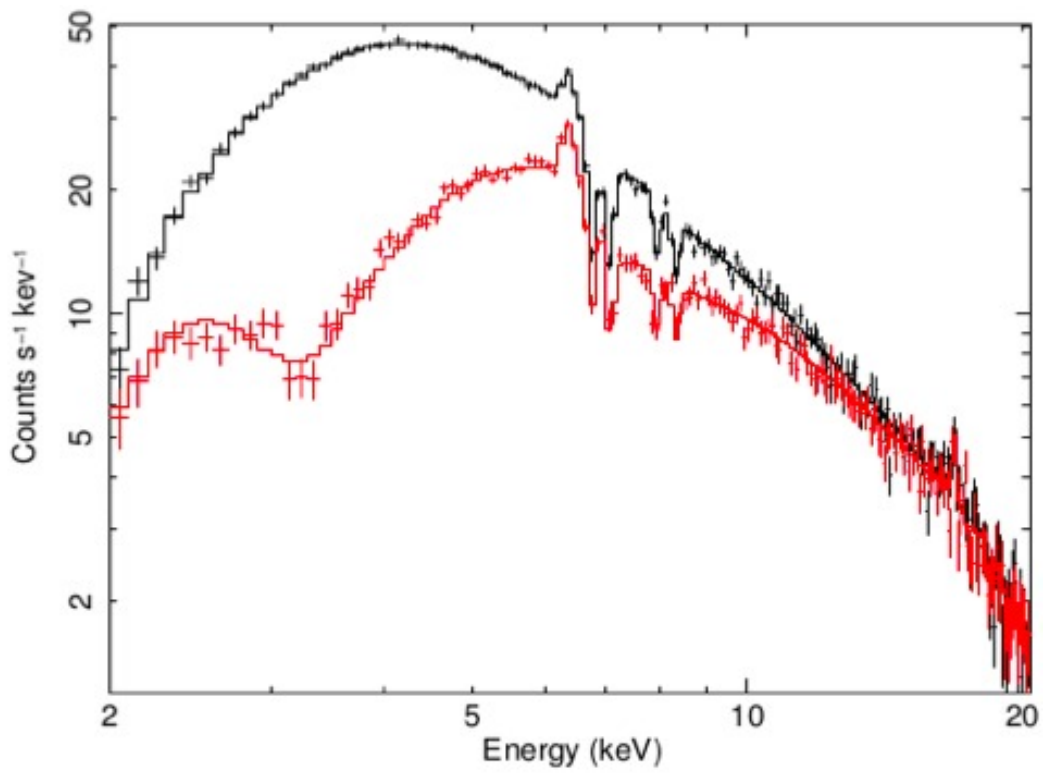

Figure 4 A simulated LAD observation of the obscured Seyfert galaxy NGC 1365, illustrating the quality of the data offered by this instrument also on relatively dim sources.

LOFT core science will take less than 50% of the mission observing time. Consequently, observatory science will be able to exploit the superb mission capabilities for more than 25 Ms in 3 years. In addition to that, the data of the WFM will be made immediately publicly available: this will provide data for >130 gamma ray burst per year, >5000 type I X-ray bursts and long-term timing and spectroscopic information on hundreds of X-ray sources of all classes.

## 3. SCIENCE REQUIREMENTS

The study of the LOFT science case resulted in the definition of a number of detailed scientific requirements for the payload and ground segment. In this section we describe only the main requirements, which drove the design of the LOFT mission.

### 3.1 Effective Area

The scientific requirement characterizing most the LOFT mission is the effective area of the LAD, the Large Area Detector, to be greater than 3.8 m$^2$, 9.5 m$^2$ and 0.8 m$^2$, at 2, 8 and 30 keV, respectively. This is achieved by deploying a detector geometric area of 15 m$^2$ (>2000 individual Silicon detectors). A number of factors affect the area effectively exposed to cosmic sources. The most significant is the open area ratio of the collimator. This causes a 30% obscuration of the detector area, energy-independent, and basically determines the peak of the effective area, in the energy range ~6-10 keV. In the baseline LAD design the effective area peaks at about 10 $m^2$. At low energies the effective area is affected by any structure intervening between the detecting medium (the Silicon) and the cosmic source, including both the external filters (e.g., the Al-coated 1µm thick polymide optical-thermal filter) and the structures on the detector itself (e.g., electrodes, implants, …). The loss in effective area increases at lower energies, as expected by the photoelectric effect, resulting in the profile shown in Figure 5. Although the nominal lower energy range for the LAD is 2 keV, the anticipated detector performance show that a lower energy threshold of 1.5 keV can be safely reached and 1.0 keV is a realistic goal. Indeed, as shown in Figure 5, the LAD effective area in the 1-2 keV region (shaded in the figure) is on average still of the order of 4 $m^2$, which provides an important contribution to the study of the softer (e.g., thermal) emission from the target sources.

At high energies (>10 keV) the factor limiting the effective area is the decreasing quantum efficiency of the 450 µm thick Silicon. As a matter of fact, the signal to noise above 30 keV becomes unfavorable also for bright sources and this determines the effective LAD upper energy bound. However, due to the increasing transparency of the capillary plate collimators at energies above ~30 keV, for bright and hard impulsive events like gamma-ray bursts and magnetar flares a large number of counts is anyway detected. For these counts, the energy reconstruction is very difficult, as they originate from photons scattered in the detector and collimator structures. However, they preserve entirely the timing features of the source and will be used to study the transient quasi-periodic oscillations in the intermediate flares from magnetar sources, which are indeed part of the LOFT core science (EOS3) and any timing feature in the prompt emission of gamma ray bursts. To enable collecting these “out-of-range” counts, the dynamic range of the LAD is expanded up to 80 keV and the counts between 30 keV and 80 keV are collected with full time resolution but with only a 2 keV energy binning (as the energy information for these counts is nearly meaningless). The limit of 80 keV was selected as a safe margin from the energy releases of the minimum ionizing particles (MIP).

The compliance with the LAD effective area requirement has been one of the driving requirements of the spacecraft design by the ESA industrial studies. The basic unit of the LAD is the Module (including 16 detectors, for an effective area of 800 $cm^2$) and the effective area requirement translates into fitting no less than 121 Modules in the LAD panels. The two spacecraft studies were indeed able to fit 124 and 125 modules, respectively, offering some margin with respect to the requirement. However, the LOFT Team also investigated the robustness of the science case against any possible loss in effective area. This study (see LOFT Yellow Book) shows that for most of the LOFT core science goals, even loss in effective area as high as 20% (which is 25 modules or 400 detectors!) or more can be compensated by longer observations of the same target. This is not applicable to only 2 of the 8 “formal” science goals, namely waveform analysis on QPO coherence timescale and Fe-line tomography in AGNs, where the intrinsic timescale of the natural events does not allow to extend the observing time beyond their duration. For these two goals, a loss in effective area would translate into a higher limiting flux (due to lower sensitivity) for these specific targets and thus to a smaller number of available targets. This analysis shows the robustness of the LOFT payload design with respect to the science case and that even a de-scoping of the mission by 20% or higher would not severely affect the LOFT mission success.

### 3.2 Spectral Resolution and Sky Visibility

A detailed analysis of the LOFT science case shows that only part of the science objectives indeed requires the full energy resolution for the LAD (240 eV FWHM @ 6 keV, end-of-life). Relaxing the energy resolution requirement for some types of observations offers the possibility to expand the flexibility of the mission. This was implemented by introducing the concepts of Field of Regard (FoR) and extended Field of Regard (eFoR). The FoR is defined as the fraction of the sky accessible by the LAD while matching all the scientific requirements (primarily, the 240 eV energy resolution). In contrast, the eFoR requires the LAD spectral resolution to be better than 400 eV, as it is meant to accommodate observations not requiring the best spectral resolution. The requirements for the FoR and eFoR are 35% and 50% at end-of-life, respectively.

Operationally, the LAD energy resolution depends on the operating temperature of its detectors which, in turn, depends on the Sun aspect angle. The operating temperature determines the value of the detector leakage current, a major contributor to the spectral resolution. This has to be defined at the end-of-life (EoL), due to the increasing radiation damage of the detectors as the orbital lifetime progresses. The requirement for the LAD detectors is <-11°C EoL.

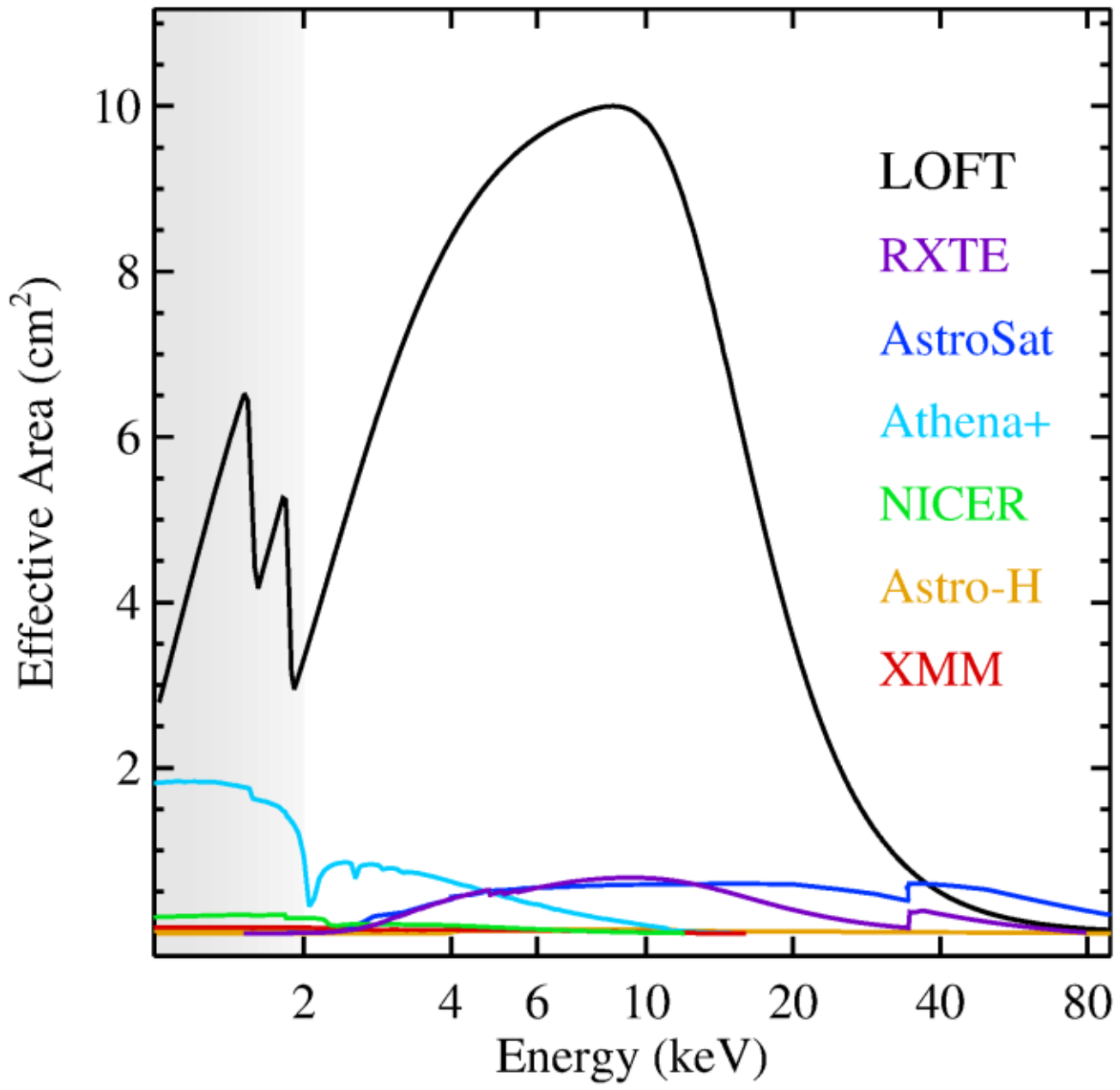

Figure 5: LOFT effective area versus energy, as compared to approved or past missions

The mission studies carried out independently by Astrium-D and Thales-I under the coordination of ESA optimized the spacecraft design to meet the detector operating temperature requirement by passive cooling on the largest possible range of Sun aspect angles. As it is shown in Figure 6, despite using two very different approaches, both spacecraft designs satisfy the requirements on the LAD detector operating temperature with large margins. This can be seen and quantified in two ways: a) the nominal energy resolution (or better) is guaranteed over a larger sky fraction than required (e.g., for the FoR, between 45% and 65%, as compared to the required 35%) or b) the energy resolution provided within the FoR is much better than the requirement (e.g., for the FoR, between 180 eV and 220 eV, as compared to the required 240 eV).

Overall, the excellent mission designs achieved by TAS-I and AST-D offer to the LOFT mission a large sky visibility: more than 75% of the sky will be accessible to LOFT at any given time of the mission lifetime (including both FoR and eFoR).

### 3.3 Mission Lifetime

The requirement on the mission duration is set by elaborating on the minimum time span that guarantees a successful mission, i.e., all the core science objectives are met. Of course, the nominal mission duration is not related to the technical duration of the payload or the ballistic life of the orbit. The real mission duration is expected to be much longer than the requirement, provided that funding for the mission operations will be secured. This analysis takes into account all the factors which decreases the observation efficiency of the mission, including maintenance, calibrations, slews and target accessibility. In fact, during the assessment study all the core science objectives have been individually analyzed, identifying which specific targets are needed and how much observing time and efficiency has to be applied to each and every of them. In combination with the larger-than-required LAD visibility, this study allowed to firmly assess the feasibility of the LOFT core science objectives in less than 3 years of mission operation. This duration also guarantees the detection of the rare transient events required for the core science. For this reason, the mission duration requirement was set at 3+2 years, implying that 3 years is the minimum operation time required by the core science, but the spacecraft and payload design and components should guarantee at least an extension for no less than additional 2 years of operation.

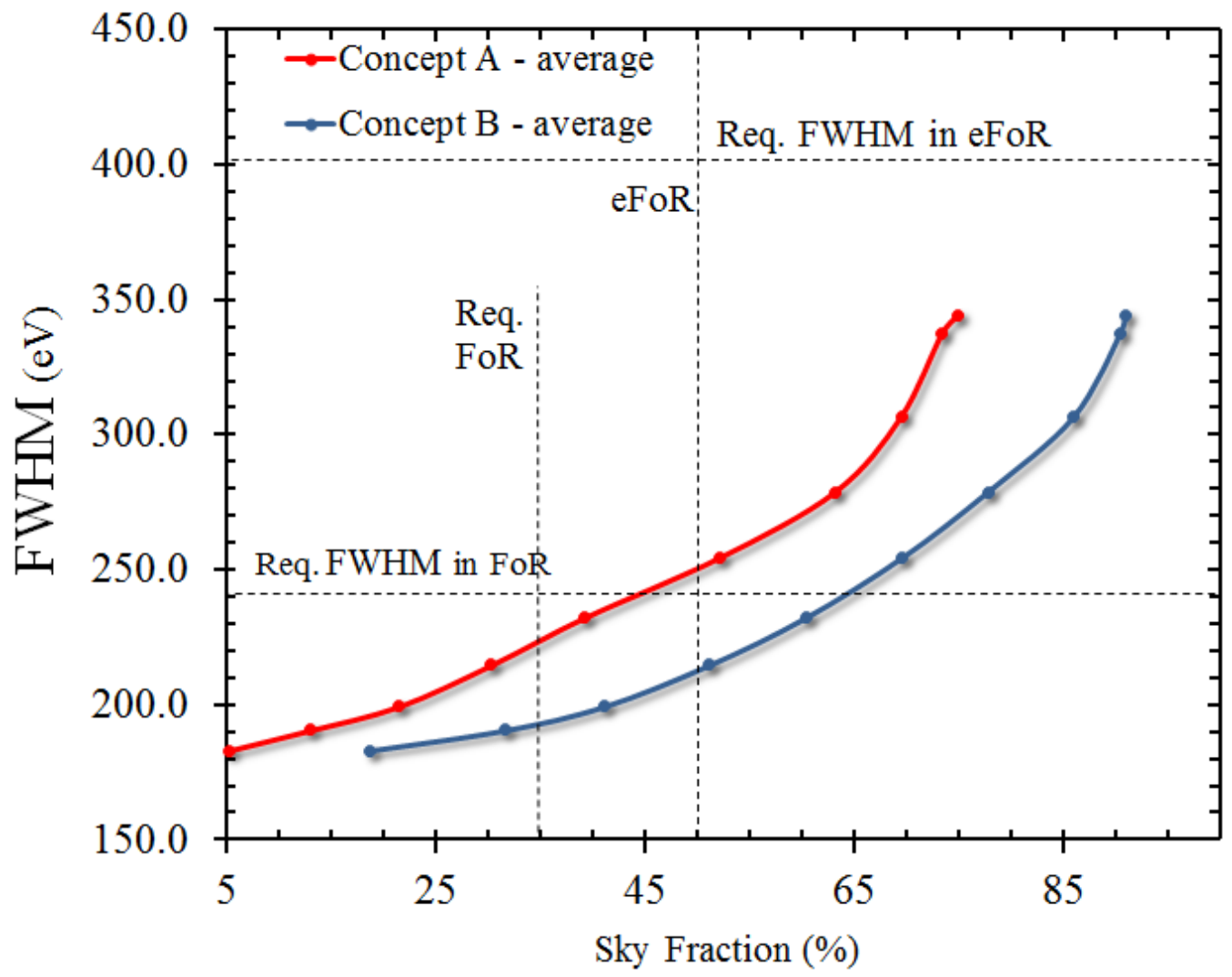

Figure 6: the results achieved by the two independent industrial studies (here anonymously indicated as "Concept A and B". Horizontal and vertical dashed lines identify the requirements of sky fraction and energy resolution for both FoR and eFoR (35% and 240 eV, 50% and 400 eV, respectively). The plot shows that both concepts are compliant with the requirements and indeed both designs provide a spectral resolution (through their operating temperature) much better than required.

### 3.4 Background

The background of the LAD experiment is a critical requirement for some of the science objectives, namely those related to the observation of dim sources (Active Galactic Nuclei), aimed at studying the variations of the 6 keV Iron line, via spectral-timing. The intrinsic scientific requirement is actually related to the amount of residual background in the source data after proper subtraction, usually referred to as background systematics. In LOFT this is controlled by two requirements: the absolute level of the background and the residual fraction, 10 mCrab and 0.25%, respectively.

The LAD background level and spectral distribution were estimated by means of a Monte Carlo simulation using the GEANT code with a detailed mass model of the satellite and instrument. The methods and the results are described in detail in Campana et al. [8]. The integrated 2-30 keV requirement of 10 mCrab is met with a 10% margin, and the spectral distribution (Figure 7) shows an optimal background (50% better than requirement) in 2-10 keV, the most important energy range from a scientific point of view. The analysis of the LAD background also shows that the "internal" component, due to particle interactions with the satellite and instrument structures, is very low (red and cyan lines in Figure 7). This is due to the very low mass of the detectors and collimators (low conversion efficiency) and the favorable equatorial orbit (low particle flux). In this respect, the LAD internal background per unit area is approximately one order of magnitude lower than a traditional large-area proportional counter array, e.g. the RossiXTE/PCA experiment. However, the stopping power of the capillary plates used as LAD collimator make the detector susceptible to a "leak" of photons with energy above ~50 keV: they are downgraded to the LAD energy range by Compton interactions with the collimator and/or the detector, resulting in valid background counts. Taking into account the Cosmic X-ray Background (CXB) and the Earth albedo, the background counts induced by these components (purple and orange lines in Figure 7) amount to a 70% of the total LAD background. Other background components include internal radioactivity (K40 decay) of the collimator glass and the aperture background (CXB photons entering the field of view).

The dominance of steady components with photonic origin (as opposed to charged particles) is a key factor in determining a very low orbital modulation of the background, which is estimated by simulations as about 10% (Figure 7, right panel, black triangles and curve), to be compared to a ~250% background modulation for massive experiment operating in a 30° inclined orbit (e.g., RossiXTE/PCA, [9]). Indeed, most of the residual 10% LAD background modulation is not due to intrinsic variations of the background sources (e.g., CXB and albedo) but to the cyclic change of their relative viewing angle by the LAD experiment along the orbit, which makes the modulation not only very limited in amplitude but also efficiently modelled and reconstructed. This is shown in the right panel of Figure 7, where the modulation of the individual components is shown, as reconstructed through a geometrical and physical model. Based on this model, a set of time-dependent simulations have been performed to estimate the residual systematics in the background subtraction, for a large set of targets and spacecraft attitudes. The results show that the systematic residual uncertainty after background subtraction is conservatively estimated as <0.15% (actually significantly better in most of the cases), to be compared to the requirement of 0.25%. This is indeed not surprising when compared to the 0.5-1.0% level reached by the RossiXTE team [10], in the far more complex conditions of a background modulation as high as 250%, largely induced by particles. With this level of background systematics, we estimated the limit flux of the LAD as 7 x $10^{-13}$ erg $cm^{-2}$ $s^{-1}$ (5 $\sigma$), reached in less than a daily exposure. This systematics do not affect the sensitivity at short timescales, which is as high as ~1 mCrab/s.

As a baseline, the LAD background model will be calibrated by periodic observations of blank fields (again, similarly to what was done by RossiXTE). As a further benchmark of the background model, a "closed module" is included in the LAD: one of the modules is equipped with a collimator which has no holes to the sky but offers the same material and stopping power as the standard modules. The data collected by this module will be available together with the standard data and will provide a continuous link between background model and real data, also allowing to monitor for any unexpected background events (e.g., solar flares).

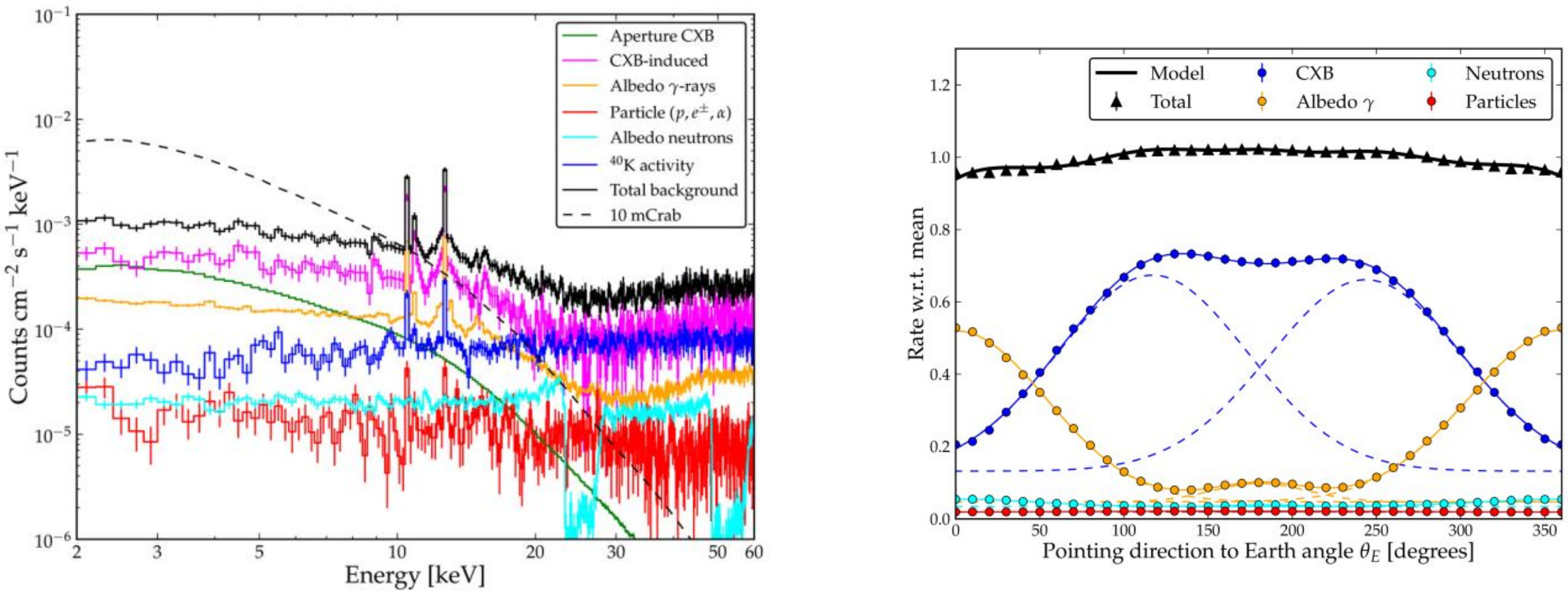

Figure 7: the background estimation for the LAD experiment, as derived by a detailed Monte Carlo simulation including a full geometrical description. *Left:* The black solid line shows the total background (sum of all components). The dashed line reports the energy-dependent requirement. *Right:* the orbital modulation of the total background rate (black line) and the individual components.

## 4. SCIENTIFIC PAYLOAD AND GROUND SEGMENT

The science objectives briefly summarized in the previous section and quantified in the scientific requirements as EOS and SFG will be addressed by a scientific payload composed of the LAD and WFM instruments and by a dedicated scientific ground segment. A more detailed description of these three subsystems may be found in Zane et al.[1], Brandt et al. [2] and Bozzo et al. [4].

### 4.1 The Large Area Detector

The LAD is the main instrument onboard LOFT. It is a 1-degree collimated instrument operating in the nominal energy range 2-30 keV (expanded to 80 keV for events outside the field of view). An enormous collecting area, peaking at 10 $m^2$ at 6-8 keV, combined with a "CCD-class" spectral resolution, required to be better than 240 eV FWHM at 6 keV, represent the breakthrough observing capabilities of the LAD. The large area is made affordable within the context of the resource budgets of a medium-class mission by its two key technologies - the large area Silicon drift detectors (SDDs) and the capillary plate collimators – enabling a very efficient effective area per unit mass/volume/power.

The large-area SDDs were originally developed [11, 12] for particle tracking in the Inner Tracking System of the ALICE experiment at the Large Hadron Collider at CERN: 1.5 $m^2$ of SDDs in ALICE have been operating successfully since 2008. In that context, they are 300 μm thick monolithic Silicon detectors with ~50 $cm^2$ effective area, read-out by two series of anodes. In the LOFT/LAD application, the SDDs are 450 μm thick, have an active area of 76 $cm^2$ each, read-out by two rows of anodes with a pitch of 970 μm. The drift channel is 35 mm long, resulting in an area of 0.3 $cm^2$ per read-out channel. The LOFT detectors are manufactured by Fondazione Bruno Kessler (Trento, Italy), under design by INFN Trieste.

The working principle of such detectors, described in detail in Rachevsky et al. [13], is as follows. An X-ray photon is absorbed in the Silicon bulk by photo-electric effect and generates a charge cloud which drifts to the collecting anodes due to an electric field, sustained by a voltage drop of ~1300V from the median plane to the edges. While drifting, the diffusion causes the charge cloud to widen, up to ~1 mm for photons absorbed near the middle plane of the detectors, 35 mm away from the anodes. The total charge may be collected by either 1 or 2 anodes, depending on the impact point of the photon. In the LAD configuration, approximately 45% of the events are collected by a single anode, while 55% are collected by two anodes: we classify them as singles or doubles, according to their multiplicity. As the signal-to-noise is related to the read-out noise of each anode, single events a fraction of 45% of the LAD area offers an energy resolution higher than the total events. In the LAD read-out architecture, this information is identified and preserved, enabling spectral studies with higher performance. The time tagging of the individual event is done at the time of its discrimination at the anode preamplifier. This implies an uncertainty related to the drift time, depending on the impact point. The maximum uncertainty is ~7 μs, for events that drift along the whole channel. The LOFT detectors have been qualified through a set of dedicated irradiation campaigns [14, 15, 16].

The technology of the capillary-plate collimator is the same as the standard microchannel plates and have already been used as collimators by the MEDA and GSPC experiments onboard the EXOSAT mission as well as currently baselined for the MIXS experiment onboard the ESA BepiColombo mission [17]. In LOFT the Pb content in the lead-glass offers enough stopping power to X-rays, making it suitable for collimating soft X-rays. The LAD configuration envisages square pores with 83 μm opening and 16 μm walls, compliant with the requirement of <1% transparency at 30 keV. The plate thickness is 5 mm, reaching the aspect ratio of 60:1 corresponding to the required field of view of 1 degree FWHM. The open area ratio is 70% and the size of each collimator tile is such to cover a whole SDD, approximately 72 mm x 112 mm. The baseline LOFT capillary plate collimators are manufactured by Photonis (Brive, France), under design of University of Leicester.

Both the detector and the collimator components in the LAD are individual tiles, ~80 $cm^2$ in area. The LAD is then intrinsically modular (and redundant). The organization of the LAD is conceptually based on $M$ independent Detector Panels, each one hosting $N$ Detector Modules, in turn composed of 16 SDDs and collimator tiles. The number of panels $M$ and modules per panel $N$ depend on the spacecraft design. In the Consortium design, $M$=6 and $N$=21, for a total of 126 Modules. The two ESA-led industrial studies envisage ($M$=2, $N$=64) and ($M$=5, $N$=25) for a total of 125 and 124 modules, respectively, providing some margin with respect to the required number of 121 modules deriving from the formal requirement of 9.5 $m^2$ peak effective area.

Each Module is equipped with its own Module Back End Electronics (MBEE) interfacing the Front End Electronics (FEEs) of the 16 detectors. Each FEE hosts 14 dedicated read-out ASICs [18]. The MBEE is in charge of providing regulated power and digital commands, as well as handling the data I/O. In the Consortium design, the 21 MBEEs of each panel are interfaced by a Panel Back End Electronics (PBEE), in turn connected with the single LAD Instrument Control Unit (ICU), including Data Handling Unit (DHU) and Power Distribution Unit (PDU). Also for the other two payload architectures the number of MBEEs served by a single PBEE remains approximately the same, 25 and 32, respectively. A detailed description of the LAD digital electronics and its architecture is given in Tenzer et al. [19].

The basic unit of the LAD is the Module. This is organized in a mechanical frame providing assembly and alignment interface to the 16 detectors and to the relevant MBEE. Each detector assembly is composed of a SDD equipped with its own FEE in a sandwich architecture where the FEE board also provide mechanical support to the SDD. Similar to the detectors, each of the 16 collimator tiles is interfaced and aligned to a common mechanical support, the collimator tray, forming a single overall grid-like structure. The detector and collimator trays are then integrated and aligned, forming a complete Module when the MBEE and the backside radiation shield and thermal radiators are included. It is worth noticing that the field of view of the LAD instrument is given by the collimators (not the SDDs) and for this reason these are the systems requiring a careful alignment. It is planned to achieve this goal through an isostatic mount of each Module in the relevant Panel structure, a technique able to provide arcsecond alignment accuracies. Additional details on the LAD mechanical layout may be found in Walton et al. [20].

The overall LAD is thus composed of ~125 Modules, for a total surface of 18 $m^2$, including 15 $m^2$ of Silicon detectors. Each of the Modules is electrically independent, although commanded through a common PBEE with the other Modules in the same panel. The total effective area of the LAD as a function of energy, as shown in Figure 5, accounts for all the factors affecting the exposed detector area (see Sect. 3). At launch, the LAD panels are stowed around the support tower to fit the rocket fairing.

The high level of segmentation (each read-out channel has an area of 0.3 $cm^2$) makes pile-up completely negligible, even when observing the ~240,000 cts/s from the Crab, which effectively translate in only 1.5 cts/s on each read-out channel. In principle, the same segmentation could also offer a completely negligible dead-time. In practice, this would require an independent handling of each channel in the read-out ASICs, largely increasing the complexity of the system. As a trade-off, we identified half of an SDD tile (114 channels) as the common read-out unit, which then drives the dead-time. With such a configuration, the dead-time during the observation of a source with a flux of 1 Crab is ~0.7%.

In Table 1 we summarize the main scientific characteristics of the LAD.

**Table 1. The main characteristics of the LAD.**

| Parameter | Value |
|---|---|
| Effective Area | 4 $m^2$ @ 2 keV<br>8 $m^2$ @ 5 keV<br>10 $m^2$ @ 8 keV<br>1 $m^2$ @ 30 keV |
| Energy Range | 2-30 keV primary<br>30-80 keV extended |
| Energy Resolution FWHM | 240 eV @ 6 keV<br>200 eV @ 6 keV (45% of area) |
| Collimated Field of View | 1 degree FWHM |
| Time Resolution | 10 µs |
| Absolute Time Accuracy | 1 µs |
| Dead Time | <0.7% @ 1 Crab |
| Background | <10 mCrab (<0.15% systematic) |
| Max Flux | 500 mCrab full event info<br>15 Crab binned mode |

## 4.2 The Wide Field Monitor

Due to the huge area and the prime scientific role of the LAD, the WFM is a “supporting instrument” in the LOFT payload, providing triggers and scientific context to the LAD observations. However, in absolute terms the LOFT WFM is the largest and most sensitive wide field X-ray monitor to date. It will monitor simultaneously 5.5 sr (at zero response, 4.1 sr at 20% of peak response), with 1 arcmin source location accuracy and 300 eV spectral resolution in the 2-50 keV energy range.

The WFM is based on the classical coded mask imaging technique. The instrument is composed of 5 Units, monitoring different but partially overlapped sky regions. Each unit is composed of 2 Cameras. As described in Brandt et al.[2], the same SDDs used in the LAD are used for imaging purposes in the WFM, by adopting a proper anode pitch (145 μm vs 970 μm of the LAD). With the same "1D" read-out electronics as the LAD, the WFM SDDs are able to localize the photon impact point with an accuracy as high as ~50-70 μm in the anode direction (by charge barycentering) and as ~3-8 mm, energy dependent, in the drift direction. The latter is achieved without any additional read-out, by measuring the width of the charge cloud reaching the anodes: the farer is the absorption point along the drift channel, the wider is the charge distribution at the anodes, due to diffusion along the drift (see [21] for details). Combining these detectors with an asymmetric coded mask (pitch 250 μm and 16.25 mm, respectively) at a distance of ~20 cm and a surrounding collimator blocking the diffuse X-ray background, the resulting camera is sensitive in the 2-50 keV, with an angular resolution of ~4 arcminute in the fine direction and ~5° in the coarse direction. For a full 2D imaging of the sources, two identical cameras, observing the same region of the sky but with the fine imaging direction rotated by 90°, form one WFM Unit. By combining the response of the two Cameras, a 2D angular resolution of 4'x4' is achieved, while keeping a strong redundancy in case of failure of a single Camera (although with coarse resolution in one direction). Individual source positions are reconstructed to better than 1 arcmin in both directions (point source location accuracy), by pixel barycentering. A discussion of the imaging properties of the LOFT/WFM may be found in Evangelista et al. [22]. Being the same as the LAD, the WFM detectors have similar intrinsic energy resolution. However, due to the finer pitch the same charge spreads over a larger number of anodes and therefore confronts a higher read-out noise. The resulting energy resolution is about 300 eV FWHM @ 6 keV (end of life, at the WFM operating temperature).

The main scope and requirement of the WFM experiment is to monitor and image >50% of the sky fraction accessible to the LAD at any time, to trigger its observations on the most interesting source states and provide a context with its long-term observation of the same sources. In coded mask experiments, the sensitivity is driven by the aperture background and vignetting factors for large off-axis angles. Large regions of the sky are then better monitored by a set of smaller units. This is indeed the strategy adopted by the WFM, as shown in Figure 8. To optimize the WFM sensitivity to weak sources, the open fraction of the coded masks has been chosen to be 25%. The resulting sky-projected WFM area is shown graphically in Figure 9.

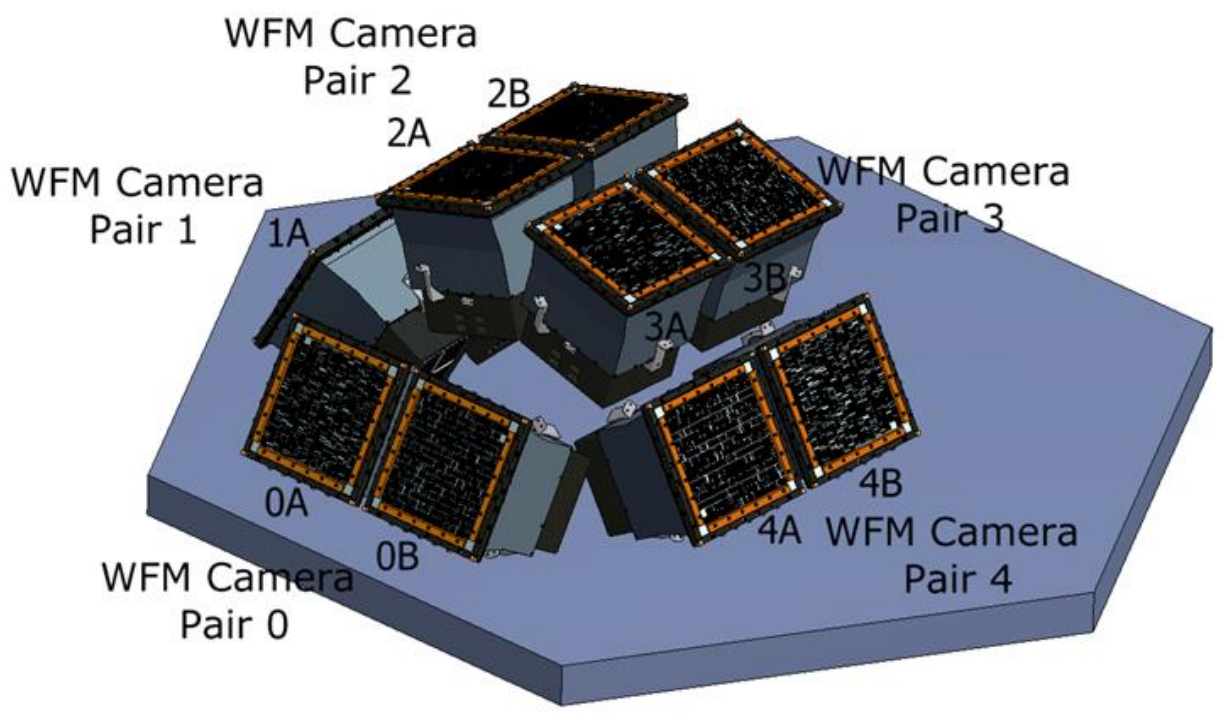

Figure 8: The layout of the 5 WFM units.

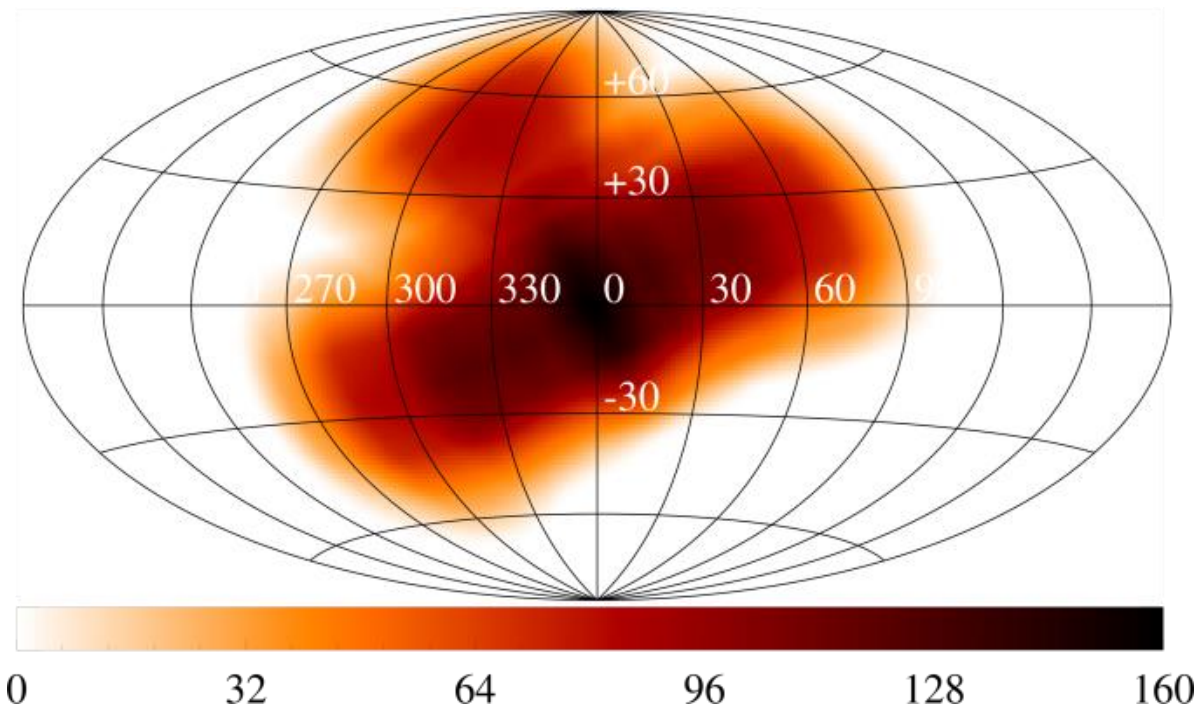

Figure 9: The projected effective area of the full WFM experiment, in Galactic coordinates.

Due to its large field of view, energy bandpass and imaging properties, the WFM is anticipated to detect and localize a large number of gamma-ray bursts (>130) and other fast transients every year. One of the WFM units is oriented in the anti-Sun direction, to facilitate follow-up observations by ground-based optical telescopes. To assist and enhance this type of observations, the onboard data processing envisages a triggering and imaging system to detect and localize transient events in nearly real-time and distribute them world-wide through a VHF transmission system. In fact, due to telemetry limitations, the WFM will normally work by integrating detector images onboard every 5 minutes (integration time is programmable), in several energy bands. Instead, for short-duration events, the onboard triggering system will enable event-by-event data storage into the mass memory, for an approximate duration of 300 seconds. The maximum sustainable rate of triggers is one per orbit, thus allowing thousands of type I X-ray bursts to be detected every year as well. In addition to its standard 100 kbits/s average telemetry allocation, the WFM will be allowed to use the telemetry bandwidth temporarily not used by the LAD allowing to transmit also full event information for (part of) the WFM standard observations.

An extensive description of the WFM experiment may be found in Brandt et al. [2]. The main scientific requirements are summarized in Table 2.

**Table 2. Anticipated performance of the WFM.**

| Parameter | Value |
|---|---|
| Energy Range | 2-50 keV |
| Active Detector Area | 1820 cm$^2$ |
| Energy Resolution FWHM | 300 eV @ 6 keV |
| Field of View (Zero Response) | 180°x90° + 90°x90° |
| Angular Resolution | 4’ x 4’ |
| Point Source Location Accuracy (10-σ) | 1’ x 1’ |
| Sensitivity (5-σ, on-axis)<br>Galactic Center, 3s<br>Galactic Center, 1 day | <br>350 mCrab<br>2.1 mCrab |
| Standard Mode | 5-min energy resolved images |
| Trigger Mode | Event-by-event (10 μs resolution)<br>Real-time downlink of transient coordinates (30s after event) |

## 4.3 The Science Ground Segment

The Science Ground Segment (SGS) for the LOFT mission follows the traditional structure of ESA astrophysics mission. The Mission Operation Center and the Science Operation Center (SOC) are under the responsibility of ESA, while the Instrument Operation Centers (IOCs) and the Science Data Center (SDC) will be provided by the LOFT Consortium. The satellite will be commanded by ESA through the ground station in Kourou, which will receive part of the data downlink, the rest being received at the Malindi ground station. The telemetry data will be downloaded every orbit (100 minutes) to the two ground stations and analyzed at SOC and SDC to identify special events, deserving a rapid reaction of LOFT or a quick dissemination to the science community. The maximum required reaction time for Target of Opportunities is 12 working hours, although based on similarly operated past missions a faster reaction is anticipated. This is consistent with the typical timescales of the transient events of interest to the core science of LOFT, which ranges from a week to several months.

Both the LAD and WFM data will be promptly analyzed in the context of a Quick Look Analysis. The LAD data will then follow the processing associated to its proprietary data rights. The WFM will instead be made publicly available to the community soon after their validation. The data policy proposed by the LOFT Consortium foresees that all LAD observing time is made available as a Guest Observer Program. The Consortium will have a minimum fraction of reserved data (order of ~20%) which will be granted on the basis of observation proposals, competing with the standard GO. If the guaranteed fraction will not come out of the standard peer review process, the relevant proposal will override the bottom of the list. This innovative approach to the data rights was conceived to guarantee that the best targets are in any case granted to the best proposers.

Due to the capabilities of the WFM to autonomously detect and localize transient events, such Gamma Ray Bursts, the LOFT payload is equipped with an onboard VHF transmitter able to distribute the relevant information to the ground within 30 s from the event. To this purpose, the LOFT ground segment includes the LBAS (LOFT Burst Alert System), which is composed of ~15 VHF ground stations, distributed over the equator to guarantee the realtime reception of the VHF alert data packet. Additional information on this system may be found in Schanne et al. [23].

Further details about the LOFT science ground segment may be found in Bozzo et al. [4].

## 5. MISSION FEATURES

The major challenge for end-of-life spectral performance of the SDDs is radiation damage induced by the trapped protons in the South-Atlantic Anomaly. This is smaller at lower inclinations and altitudes. For this reason an orbit with low-inclination (<2°), low-altitude (~550 km) has been selected. This is also very favorable from the point of view of the particle-induced instrumental background, although this is not the dominant component in the LAD.

The attitude and orbit control system (AOCS) for LOFT is required to be 3-axis stabilized. However, the huge counting statistics provided by LOFT on bright sources will require an accurate control on the possible systematic uncertainties introduced by the LAD response stability, as due to the AOCS or other sources of instability (e.g., thermal). This issue has been studied in detail by the LOFT Science Team as a function of the frequency, taking into account the expected astrophysical signal in each frequency range, as well as the discovery space open by the LAD counting statistics. The resulting requirement (rms) goes from 2% stability (per decade) below 0.01 Hz, to 0.2% in 0.01-1 Hz, to 0.02% (per octave) in 1 Hz – 1 kHz. The approach to meeting such a requirement is to combine the angular response of the collimator (ideally triangular, but in practice smoothed by the element-to-element misalignment) with the AOCS parameters: the absolute pointing error should be such that (~arcminute) the source is always in the most flat, central part of the LAD field of view, whereas the relative pointing error should be small enough (a few to 15 arcseconds, depending on frequency) that the target source “samples” a region of the LAD response varying less than the required amplitude stability, for the given frequency range. It is important to note that such requirements do not directly apply to the AOCS parameter, since the attitude control frequencies will be significantly smoothed by the satellite structure, before being transferred to the LAD panels.

The total observing time needed to satisfy all the top-level science goals has been calculated on the basis of a detailed analysis of the required and available targets and statistical quality of the observation. The result is that 24.7 Ms of net observing time are required for the Core Science, of which 18.8 Ms also require the nominal spectral resolution. The mission availability (thus excluding maintenance, Earth occultation, slews, safe modes and any temporary loss of observing efficiency) has been estimated by the ESA contractors as 60%. A mission duration of 3 years offer ~57 Ms of net observing time. This leaves more than 60% of the observing time available to the Observatory Science targets, in addition to provide ample margin to the allocation of core science observations. It should be noted that the 3 years mission duration is also related to guarantee a high enough probability to catch rare transient events which are part of the LOFT’s primary targets.

The main mission features are listed in Table 3.

**Table 3 The main characteristics of the LOFT mission.**

| Parameter | Value |
|---|---|
| Orbit | Equatorial, 550 km |
| Launcher | Soyuz (6,000 kg launch capability) |
| Mass | 4,000 kg |
| Power | 4 kW |
| Telemetry | 6.7 Gbit/orbit |
| Ground Stations | Kourou, Malindi – X-band |
| Pointing | 3-axis stabilized |
| Mission Duration | 3+2 years |

## 6. CONCLUSIONS

The Assessment Study of the LOFT ESA M3 mission candidate has been carried out by a consortium of European institutes for the payload instrumentation, and by the ESA Study Team and its industrial contractors for the spacecraft and system aspects. The payload study by the LOFT Consortium as well as the spacecraft studies by the Astrium-D and Thales-I were reviewed by an ESA panel at the Preliminary Requirements Review (PRR). This review confirmed the feasibility and low risk of the mission within the programmatic boundary conditions of the ESA M3 mission. Also the payload technologies were evaluated as mature and low risk. Launch in 2023 was evaluated as realistic (the M3 launch date is scheduled on 2024). The results of the study are reported in the technical and scientific reports of the Assessment Study. The latter is publicly available in the LOFT Yellow Book on the ESA web site[1].

Following the PRR the LOFT mission participated into the final M3 mission selection by the Advisory Structure and Science Program Committee of ESA in February 2014. The LOFT science case was evaluated as very strong and the payload and observing strategy as solid, enabling the achievement of the science goals. LOFT was acknowledged as better suited than any other planned ESA or international X-ray facilities to address the science case of the equation of state of ultradense matter and the strong field gravity. However, despite a very positive evaluation, LOFT was not selected as M3 mission.

Considering the international scenario, the science case of LOFT remains unique and not adequately addressed by other planned projects. For this reason, the LOFT Consortium will continue its efforts on this mission and will propose it for the next flight opportunities.

## ACKNOWLEDGEMENTS

The Italian team is grateful for support by ASI (under contract I/021/12/0-186/12), INAF (under contract PRIN-INAF-2011 “Strong Gravity”) and INFN (under contracts XDXL and REDSOX). The work of the MSSL and Leicester groups is supported by the UK Space Agency. The work of SRON is funded by the Dutch national science foundation (NWO).

[1] http://sci.esa.int/loft/53447-loft-yellow-book/

The work of the group at the University of Geneva is supported by the Swiss Space Office. The work of IAAT on LOFT is supported by the Germany's national research center for aeronautics and space DRL. The work of the IRAP group is supported by the French Space Agency. LOFT work at ECAP is supported by DLR under grant number 50 00 1111. IEEC-CSIC has been supported by the Spanish MINECO (under grant AYA2011-24704).

## REFERENCES

[1] S. Zane et al., this volume (2014).
[2] S. Brandt et al., this volume (2014).
[3] M. Feroci et al., "The Large Observatory for X-ray Timing (LOFT)," Experimental Astronomy, 34, 415 (2012).
[4] E. Bozzo et al., this volume (2014).
[5] K.H. Lo et al., Astrophysical Journal, 776, 19 (2013).
[6] A. Watts et al., in preparation (2014).
[7] P. Uttley et al., Astron. And Astrophysics Review, in press (2014).
[8] R. Campana, et al., Exp. Astronomy, 36, 451 (2013).
[9] K. Jahoda et al., "Calibration of the Rossi X-ray Timing Explorer Proportional Counter Array," ApJS, 163, 401 (2006).
[10] N. Shaposhinikov et al., Astrophysical Journal, 757,159 (2012).
[11] A. Vacchi et al., "Performance of the UA6 large-area silicon drift chamber prototype," Nuclear Inst. and Methods in Physics Research A, 306, 187 (1991).
[12] A. Rashevski et al., "Large Area Silicon Drift Detector for the Alice Experiment," Nuclear Inst. and Methods in Physics Research A, 485, 54 (2002).
[13] A. Rachevski et al., Journal of Instrumentation, submitted (2014).
[14] E. Del Monte et al., this volume (2014a).
[15] E. Del Monte et al., Journal of Instrumentation, in press (2014b).
[16] E. Perinati et al., this volume (2014).
[17] G.W. Fraser et al., "The mercury imaging X-ray spectrometer (MIXS) on BepiColombo" , Planetary and Space Science, 58, 79 (2010).
[18] A. Cros et al., this volume (2014).
[19] C. Tenzer et al., this volume (2014).
[20] D. Walton et al., this volume (2014).
[21] R. Campana et al., "Imaging Performance of a large area Silicon Drift Detector for X-ray astronomy," Nuclear Inst. and Methods in Physics Research A, 633, 22 (2011).
[22] Y. Evangelista et al., this volume (2014).
[23] S. Schanne et al., this volume (2014).